\documentclass[aps,twocolumn,showpacs,preprintnumbers,pra,10pt]{revtex4-1}
\usepackage{graphicx,dcolumn,bm,amsmath,amssymb,hyperref,xcolor}
\usepackage[english]{babel}

\newcommand{\n}[1]{{#1}}
\newcommand{\delete}[1]{{ }}
\newcommand{\elp}{{ \ell_P }}
\renewcommand{\a}{\alpha}

\renewcommand{\b}{\beta}
\renewcommand{\d}{\delta}
\newcommand{\eps}{\epsilon}

\newcommand{\D}{\Delta}
\newcommand{\expv}[1]{\langle #1 \rangle}
\newcommand{\crr}{C_{\rho\rho}}
\newcommand{\ccrr}{C^{\textrm{nl}}_{\rho\rho}}

\newcommand{\br}{\bm{r}}

\newcommand{\brp}{\bm{r}'}

\newcommand{\Ec}{E_c}

\begin{document}
\title{Length scale of puddle formation in compensation-doped semiconductors and topological insulators}
\author{\n{Thomas  B\"omerich}}
\author{Jonathan Lux}
\author{Qingyufei Terenz Feng}
\author{Achim Rosch}
\email{rosch@thp.uni-koeln.de}
\affiliation{Institute for Theoretical Physics, University of Cologne, D-50937 Cologne, Germany}
\date{\today}

\begin{abstract}
\noindent
In most semiconductors and insulators the presence of a small density of charged impurities cannot be avoided, but their effect can be reduced by compensation doping, i.e. by introducing defects of opposite charge. Screening in such a system leads to the formation of electron-hole puddles, which dominate bulk transport, as first recognized by Efros and Shklovskii. \n{Metallic surface states of topological insulators (TI) contribute extra screening channels, suppressing puddles.
We investigate the typical length $\elp$, which determines the distance between puddles and the suppression of puddle formation close to metallic surfaces in the limit where the gap $\Delta$ is much larger than the typical Coulomb energy $E_c$ of neighboring dopants, $\Delta \gg E_c$. 
In particular, this is relevant for three dimensional Bi-based topological insulators, where $\Delta/E_c \sim 100$. Scaling arguments predict $\elp \sim (\D/E_c)^2$. In contrast, we find numerically that $\elp$ is much smaller and grows in an extended crossover regime approximately linearly with $\D/E_c$ for numerically accessible values, $\D/E_c \lesssim 35$. We show how a quantitative scaling argument can be used to extrapolate to larger $\Delta/E_c$, where  $\elp \sim (\D/E_c)^2/\ln(\D/E_c)$. Our results can be used to predict a characteristic thickness of TI thin films, below which the sample quality is strongly enhanced.}
\end{abstract}

\maketitle
\section{Introduction}
The formation of locally conducting puddles is a phenomenon caused by charged Coulomb disorder in insulators, semiconductors and Dirac-matter like graphene, topological surface states or Weyl semimetals.
Efros and Shklovskii \cite{ESbook} predicted that puddle formation is, in three dimensions, an unavoidable consequence of the long-range nature of the Coulomb interaction. Puddles are formed to screen large potential fluctuations exceeding the size of the gap $\Delta$.

In graphene it has been shown both theoretically and experimentally that puddles are necessary to understand most transport experiments 
close to charge neutrality
\cite{Hwang2007, Adam2007,Martin2008,Zhang2009}. Recently, they have been observed for the first time in the bulk of a three-dimensional topological insulator \cite{Borgwardt2016, Rischau2016}.
These materials, from the class of the Bi$_{2-x}$Sb$_x$Te$_{3-y}$Se$_y$ compounds \cite{Ren2011}, are almost perfectly compensated semiconductors with a band gap of order $250-300$ meV almost 2 orders of magnitude larger than the typical Coulomb energy $E_c$ of neighboring dopants  \cite{Borgwardt2016}.
The relatively high density ($>10^{19} \rm{cm}^{-3}$) of dopants implies a strongly fluctuating Coulomb potential in the bulk. This leads to band bending and eventually to the formation of electron and hole puddles \cite{Skinner2012, Skinner2013Rev}.
The additional surface states in the topological materials induce an additional screening channel close to the surface. Here surface puddles form \cite{Skinner2013Rev,Skinner2013} which are akin to puddles that form in graphene on a substrate
which has charged impurities.

As the puddles are separated by insulating regions, they do not directly contribute to the DC conductivity.
However, they do contribute to the optical conductivity at finite frequencies, which has been used to detect their presence and to measure the effective charge density in conducting regions \cite{Borgwardt2016}. 
We have shown that screening from thermal excitations can efficiently suppress puddle formation leading to a characteristic temperature dependence of the optical response \cite{Borgwardt2016}.
Furthermore, in similar compounds a giant negative magnetoresistance was found experimentally and explained by merging of puddles driven by the Zeeman effect \cite{NMR}.

Surface puddles and puddles in graphene can be observed directly in real space by scanning tunneling microscopy (STM) \cite{Martin2008, Zhang2009, STM2011}. 
From the two-dimensional STM maps, the size of the potential fluctuations and the corresponding length scale can be directly read off.
These agree well with theoretical results, where these quantities are calculated self-consistently \cite{Hwang2007, Adam2007, Skinner2013}.
However, nothing is known experimentally about the length scales of puddles in the bulk and the effect of surface screening on the bulk puddle formation.

In the following we demonstrate numerically that the length scales governing \n{the distance of puddles, the suppression of  (bulk) puddles close to surfaces of TIs, and the suppression of puddles in thin films  grow much slower with $\D/E_c$ than expected from scaling arguments.}
First we introduce the model and consider the scaling behavior of the charge-charge correlation function. 
We show numerical results for the bulk, and demonstrate that the simple scaling theory fails.
Then we additionally take into account the gapless surface states which provide an extra screening channel. 
The length scales governing the size of puddles on the surface is different, and independent of the bulk band gap \cite{Skinner2013}. 
The bulk length describes, however, the size of a region where surface screening suppresses the formation of bulk puddles and is therefore important to understand the properties of thin topological insulator samples. \n{We use scaling arguments to extrapolate our numerical results for $\Delta/E_c\lesssim 35$ to the experimentally relevant regime of $\Delta/E_c\sim 100$.}

\section{Model and Simulations}\label{model}
Bi-based topological insulators typically have a very large dielectric constant $\varepsilon \approx 200$. Electron binding energies are therefore small.
Thus, the bare energies of the dopants are located very close to the band edges and can be approximated by $+\Delta/2$ for the donors and $-\Delta/2$ for the acceptors.
To model the non-linear screening of randomly placed charged impurities in such a system we use a simple classical model \cite{Basylko2000, Skinner2012, Skinner2013Rev,Borgwardt2016}:
\begin{equation}\label{eq:ham_units}
 H = H_n+H_C=\frac{\Delta}{2} \, \sum_i f_i n_i +\frac{1}{2} \sum_{i \neq j} V_{ij} \; q_i q_j
\end{equation}
where $f_i=\pm 1$ are random numbers with
$f_i=+1$ for a donor states and $f_i=-1$ for the acceptor state at position $\br_i$. $V_{ij}$ denotes the Coulomb interaction between the dopants at positions $\br_i$ and $\br_j$.
$n_i \in \{0,1\}$ denotes the electronic occupation of the $i-$th dopant and is determined by minimizing the Hamiltonian. It is related to its charge $q_i$ (in units of $|e|$ where $e$ is  the electron charge) by
\begin{equation}\label{eq:chargedef}
 q_i = \frac{f_i+1}{2}-n_i.
\end{equation}
A donor (acceptor) in its ground state is characterized by $f_i=1$, $n_i=0$ and $q_i=1$ ($f_i=-1$, $n_i=1$, $q_i=-1$). Somewhat counter-intuitively, screening occurs when the Coulomb interaction drives donors or acceptors into a neutral states with $q_i=0$. Several neutral donor states close by form an electron puddle, while neighboring neutral acceptor states form hole puddles.
The Coulomb energy is modeled by
\begin{equation}\label{eq:Vcutoff}
V_{ij}=\frac{e^2}{4 \pi \varepsilon \varepsilon_0 \sqrt{|\bm{r}_i-\bm{r}_j|^2+a_B^2}}= \frac{E_c}{\sqrt{|\bm{x}_i-\bm{x}_j|^2+1^2}}.
\end{equation}
Here the short-distance cutoff $a_B=\tfrac{4 \pi \varepsilon_0 \varepsilon}{m^* e^2}$ was introduced by
Skinner {\it et.~al.~}\cite{Skinner2012, Skinner2013Rev} to take into account that the wave function of the bound state is smeared over a length scale set by the effective Bohr radius of the impurity state.  Skinner {\it et.~al.~}\cite{Skinner2012, Skinner2013Rev}  argued that due to the large dielectric constant in Bi based topological insulators, $a_B$ is large and of similar size as the typical distance of dopants. We  use $a_B=N^{-1/3}$ where $N=N_A=N_D$ is the density of  dopants 
where we assume a perfectly compensated system where the density of donors equals the density of acceptors $N_A=N_D$.  For the last equality in Eq.~\eqref{eq:Vcutoff} we expressed all distance in units of the average dopant distance $N^{-1/3}$. Here 
\begin{equation}\label{Ec}
 E_c=\frac{e^2 N^{1/3}}{4\pi \varepsilon \varepsilon_0}
\end{equation}
is the typical energy scale describing the Coulomb interaction of neighboring dopants.
A large $\varepsilon\sim 200$ leads to a small energy scale $\Ec \sim 3.3 \,\rm{meV} \sim 40 \,K$, (assuming a typical density $N=10^{20} \, \rm{cm}^{-3}$) about 2 orders of magnitude smaller than
typical band gaps $\Delta$. Indeed, in Ref.~\cite{Borgwardt2016} we used the temperature dependence of the optical response to determine $E_c$ and found $\Delta/E_c \approx 150$, \n{similar parameters have also been found in Ref.~\cite{NMR}}. In the following we assume $T \ll  E_c \ll \Delta$ and consider therefore only properties at $T=0$.

The model \eqref{eq:ham_units} describes how donor and acceptor states interact with each other. It does not include the states in the electronic bands. 
This turns out \cite{Borgwardt2016} to be well justified in the limit $\Delta /E_c \gg 1$  as the density of the relevant electronic states is much smaller than the density of dopants in this limit.

To find the true ground state of the model in Eq.~(\ref{eq:ham_units}) is an exponentially hard problem, but there is an algorithm to find an approximate ground state, called a pseudo ground state, in polynomial time \cite{ESbook, Skinner2012}.
The physical properties of a pseudo ground state are expected to be indistinguishable from that of the true ground state. The single particle energies are defined as
\begin{equation}\label{eq:singleparten}
 \epsilon_j = \frac{\Delta}{2} f_j - \phi_j = \frac{\Delta}{2} f_j - \sum_{i \neq j} V_{i j} \; q_i.
\end{equation}
In a pseudo ground state
\begin{equation}\label{eq:ESstable2}
\D E_{(\a,\b)} = \eps_\b - \eps_\a - V_{\a \b}>0
\end{equation}
has to be fulfilled for all 
 pairs with $n_\b =0, n_\a = 1$. This state can be reached by exchanging electrons between states where this condition is not met. The algorithm is described in detail in Refs.~\cite{Skinner2012, Skinner2013Rev}.
Simulations are performed in a cubic volume $V=L^3$ with periodic boundary conditions with
$2 L^3$ dopants, typically we use $L=50$ or $L=60$ corresponding to  $250000$ or $432000$  
dopants. 
Numerical results shown below are averaged over $200-800$ disorder realizations, i.~e.~random configurations of the dopant positions.
We have checked \cite{LuxThesis} that our code reproduces published results (e.g. on the Coulomb gap in the density of states) from other groups \cite{Skinner2012}  on the same model in all quantitative details.
In the following we use dimensionless units where all length are measured in units of $N^{-1/3}$, and all energies are measured in units of $\Ec$ defined in Eq.~(\ref{Ec}). In these units the only free parameter of our model is $\Delta$ besides the (dimensionless) system sizes considered in Sec.~\ref{surface}.

\section{Length scales and scaling}\label{scaling}
\begin{figure}[tb]
	\centering
		\includegraphics[width=\columnwidth]{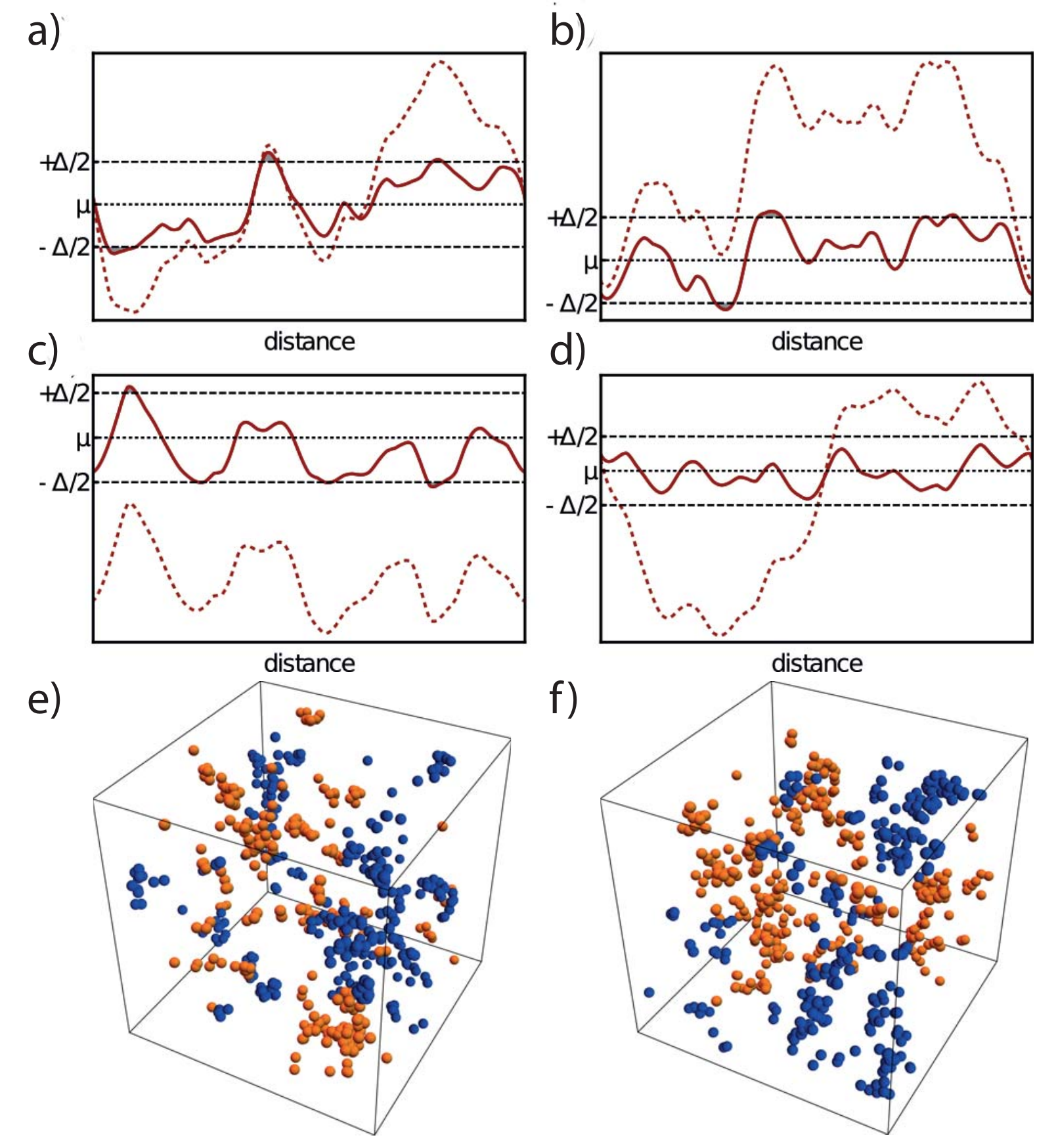}
	\caption{(Color online)  Due to charged impurities, the potential fluctuates in space. Huge fluctuations of the potential in the uncorrelated state (dashed line), where
	all dopants are charged, are screened by the formation of electron-hole puddles. The potential  $\phi(\br)$   (solid line, panels a)-d)) obtained in the ground state is restricted to the range $[- \D/2-\Ec,\D/2+\Ec ]$.
	Puddle formation occurs in tiny regions (gray shading) where $\phi(\br)$ exceeds the band edges and is thus above $\D/2$ or below $-\D/2$. Panel a)-d) shows four one-dimensional cuts through the three-dimensional potential. \n{Panel e) and f) display two examples of three-dimensional configurations of electron (orange) and hole (blue) puddles. Only the neutral dopants within the puddles are shown (about 2\% of the total number of dopants). The plots suggest that more than one length scale governs puddle formation, see text. } Plots are taken from simulations with $\D=10$, periodic boundary conditions and $L=50$ (250000 dopants), the plots show boxes of size 25.}	
	\label{potfluct}
\end{figure} 
One of the main questions that we will address is the following: What is the typical distance between electron and hole puddles or, equivalently, 
on what length scale does the potential typically change by an amount of $\D$?
It turns out that this length scale \n{also characterizes screening properties on average, as discussed in more detail below.}

A simple scaling argument by Efros and Shklovskii \cite{Efros1972} suggests
that the corresponding length scales as $R_g\sim\Delta^2$.
The argument is as follows: in a volume of size $V \sim R^3$ there are on average $N R^3$ positive and negative charges where $N$ is the density of dopants. But these two numbers are not exactly equal, instead the typical charge of the region is  (in the uncorrelated state)  $Q_R \sim \pm \sqrt{N R^{3}}$. This implies a typical potential 
of order $\phi_R \sim Q_R/R \sim \sqrt{R}$ within that region.
The fact that this potential diverges for $R\to \infty$ shows that this situation is unstable and 
the huge potential fluctuations have to be screened.
The potential can be screened when the Coulomb potential is sufficiently strong to change the charging state of the dopant. This is possible for
 $ \phi \sim \pm \D/2$. Using that $ \phi\sim \sqrt{R}$, this strongly suggests that the typical length scale $R_g$, describing both the screening length and the length scale where the potential changes by $\pm \Delta$, is proportional to $\Delta^2$.
Accordingly, the typical charge density in a volume $V=R_g^3$ is $\rho_g \sim Q_{R_g}/V \sim \sqrt{R_g^3}/R_g^3=R_g^{-3/2} \sim 1/\D^3$.
To summarize, this scaling argument suggests
\begin{equation} \label{eq:ESscaling}
 \centering R_g \sim \D^2 \textrm{ and } \rho_g \sim \D^{-3}.
\end{equation}
Restoring dimensionfull units, these equations read $R_g \sim N^{-1/3} (\Delta/E_c)^2$ and $\rho_g \sim \pm e N (E_c/\D)^{3}$. 
\n{We will show below that our numerical results for $\D/E_c \lesssim 35$ show a much slower growth of length scales with $\Delta/E_c$. We will attribute this to a huge crossover regime and the presence of logarithmic corrections obtained from a refined version of the scaling argument in section \ref{surface}.}

\delete{The length $\sim$ energy$^2$ scaling of Eq.~\eqref{eq:ESscaling} arises technically from
the properties of the uncorrelated state where charged dopants are randomly placed. 
For such a model, one can calculate exactly the probability distribution of the potential as function of system size $L$ and one finds, for periodic boundary conditions, $P_{\rm uc} (\phi) = {\rm exp} (-\tfrac{\phi^2}{2 a L})/\sqrt{2\pi a L}$ with  $a \approx 4$.
This implies that the typical potential is of order $\sqrt{L}$.
The argument above implicitly assumes that this simple random walk-like scaling is carried over to the correlated ground state despite the fact that screening leads to a `correlated' redistribution of charges.}

In Fig.~\ref{potfluct} we compare several one-dimensional cuts of the potential in the uncorrelated state (dashed lines) and in the correlated ground state (solid lines) obtained from numerical simulations.
The potential fluctuations of the uncorrelated state are much larger than $\pm \Delta/2$ (shown here for $L=50$) triggering screening. 
For the correlated ground state, in contrast, the potential fluctuations are strongly reduced and lie within the band gap. 
Puddles are formed in the tiny regions (shaded in gray) where $|\phi|$ slightly exceeds $\D/2$. One finds that in these regions $|\phi|-\D/2 \sim E_c$. 

\delete{While the cut in panel a)  shows  the expected behavior, the cuts in panels b), c) and d) show more complex phenomena. In panel b), for example, the renormalized potential changes directly from the hole- to the electron puddle on a short distance without the type of fluctuations expected from a random-walk process implicitly assumed in the scaling analysis. Panel c) shows that there is a subtle interplay of physics at large and small length scales while panel d) shows that even large fluctuations of the uncorrelated potential cannot predict puddle formation in the correlated ground state. }\n{The $3d$ plots in panel e) and f) show directly the puddles. Neutral donors constitute electron puddles and are colored in orange, while neutral acceptor are part of hole puddles, colored in blue.
The snapshots and the cuts suggest that not only a single, but several length scales govern puddle formation  \cite{Bara1984,Lee1999,Skinner2016}. The short one, $\elp$ governs the closest distance of puddles and rapid fluctuations of the potential as shown in panel b). Much longer length scales govern the formation of lengthy, anisotropic cluster structures and also regions without puddles, see panel d). }
\delete{
All these examples show that the simple picture on which the scaling argument is based, may not be able to capture the correlated state correctly. }

To obtain more quantitative results, one can study the statistical properties of either the potential $\phi(\br)$ or
directly of the charge distribution $\rho(\br)$, since both are related by the Poisson equation $ \nabla^2 \phi = -\rho$ (up to the short-distance cutoff $a_B$ introduced above).
In the following, we will mainly discuss the charge-charge correlation function $\crr$. We split this into a local part $\sim\delta(\br-\brp)$ and a non-local part $\ccrr$:
\begin{equation} \label{eq:rhorhoparam}
 \crr(\br,\brp) = \langle \rho(\br)\rho(\brp) \rangle = Q_0 \delta(\br-\brp)+ \ccrr(\br-\brp),
\end{equation}
where we used charge neutrality $\langle \rho \rangle =0 $.
Here and in the following the expectation value $\expv{\cdot}$ denotes a disorder average. After disorder averaging all correlation functions only depend on the distance $r=|\br-\brp|$.
Thanks to charge neutrality we know that $\int d^3 \br \, \ccrr(\br) = -Q_0$.
The weight of the $\d$-peak $Q_0$ corresponds to $2 N (1-n_0)$ where $n_0$ is the fraction of neutral dopants.

\section{Screening  in the bulk} \label{bulk}
\begin{figure}[tb]
	\centering
		\includegraphics[width=\columnwidth]{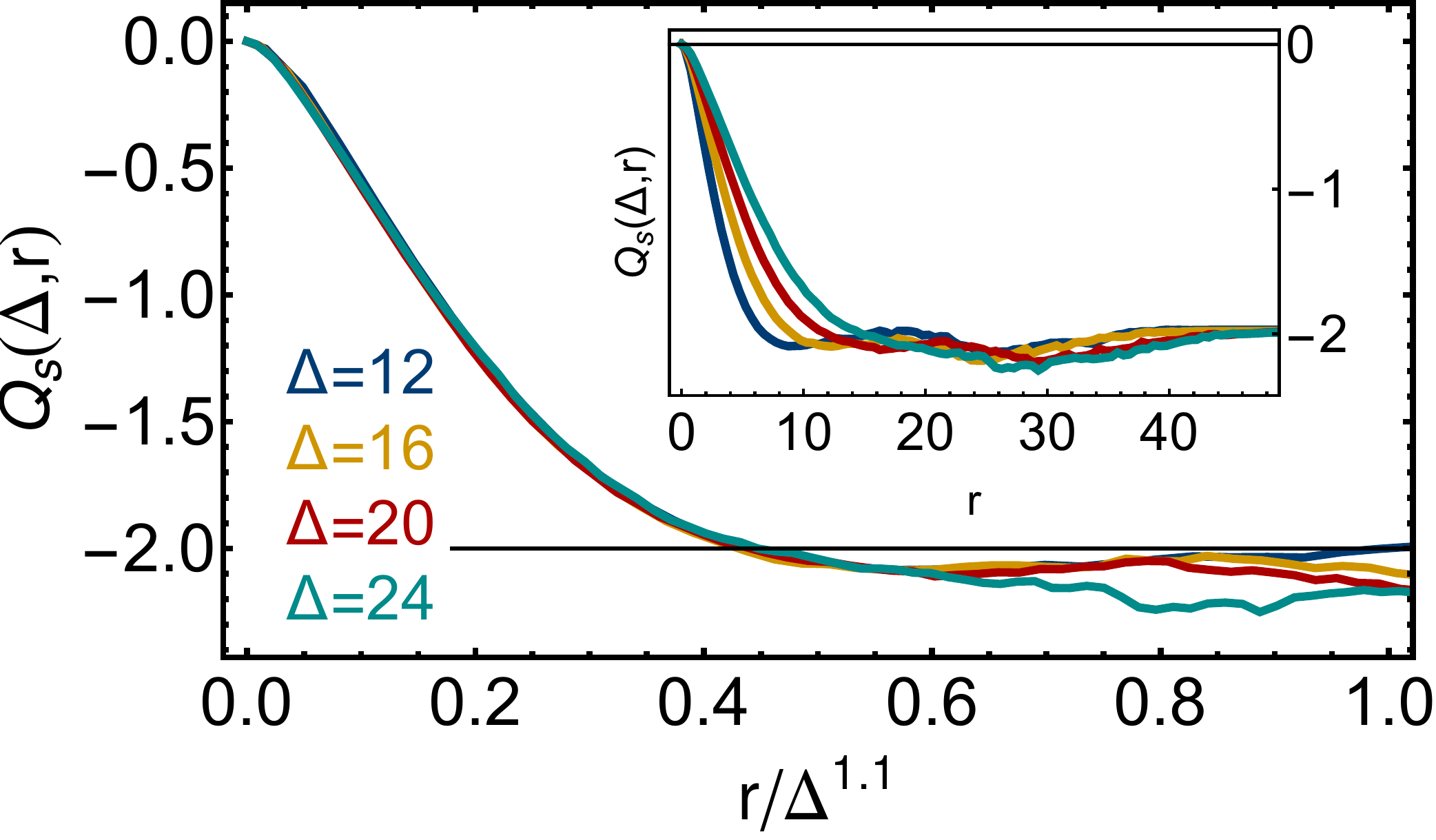}
	\caption{(Color online) \n{Apparent} scaling of the screening charge defined in Eq.~(\ref{eq:srr}) for different values of the band gap $\Delta$. The best scaling collapse is found for an exponent $\gamma=1.1$ characterizing an extended crossover regime.
	The inset shows the unscaled data.
	Deviations from the scaling behavior can be seen for $r>0.6\,\D^{1.1}$. Parameters are $L=50$ (250000 dopants) for $\D=12,16$ and $L=60$ (432000 dopants) for $\D=20,24$, and we checked 
	that there are no significant finite size effects.}	
	\label{bulksc}
\end{figure}

\n{Screening in insulating charged Coulomb systems is a highly non-local and non-linear mechanism. Early work by Baranovskii, Shklovskii and Efros \cite{Bara1984} (see also a lucid discussion in  Ref. \cite{Lee1999}) pointed out that adding a single charge can trigger an avalanche of 
discrete changes of the charge of dopants not only in the neighborhood of the charge but also at large distances. This appears to be a highly anisotropic, non-local (and perhaps fractal \cite{Skinner2016}) process. The change of the potential at larger distances is random in sign but does not decay rapidly. In contrast to a metal, there is therefore no true screening (as is also obvious from the fact 
that the system is characterized by a Coulomb gap). In the following we will not track these changes  but focus on the shorter length scale $\elp$ which governs the impurity-averaged charge correlations $\ccrr(\D,s)$, but also controls the typical `nearest' distance of oppositely charged puddles. Later we will argue that the same length scale also governs the impact of metallic surface states on puddle formation.
 }

Instead of studying directly the charge-charge correlation function $\ccrr(\D,s)$, we find it more convenient to investigate the distance dependence of the `screening charge' defined by
\begin{equation} \label{eq:srr}
Q_s(\Delta, r) = 4 \pi \int_0^{r} ds\,s^2\, \ccrr(\Delta, s).
\end{equation}
The advantage of this quantity is that it has a direct physical interpretation: it describes the charge accumulated \n{-- on average --} around a dopant  within the radius $r$ multiplied with the charge of that dopant and the density of dopants.  As negative charges accumulate around a positive charge and vice versa, the screening charge is always negative.
 Total charge neutrality requires that around a positive (negative) charge exactly the charge $-1$ ($+1$) accumulates for $r\to \infty$. As neutral dopants do not contribute, one therefore obtains
$ Q_s(\Delta, r \to \infty)=-2 N (1-n_0)=-Q_0$. This also follows directly by integrating Eq.~(\ref{eq:rhorhoparam}) over $\br$ in a charge-neutral system.
In our simulations we use boxes of size $L$ with periodic boundary conditions. For $r>L/2$ we therefore have to replace in the integral in Eq.~(\ref{eq:srr}) the factor $4 \pi s^2$ 
by $W(s)=\int \delta(s-|r|) d^3 r$. This does not affect the scaling plots discussed below but is useful to check for overall charge neutrality.

We show numerical results for $Q_s$ in Fig.~(\ref{bulksc}). On a rather short length scale (see inset) the screening charge reaches the value $-Q_0 =-2 N(1-n_0)\approx -2$ (the plot uses units where $N=1$ and
the fraction of neutral dopants, $n_0$, is less than $2\%$ for all shown values of $\Delta$).
The scaling plot (main figure) suggests that the length scale $\elp$, on which the screening occurs, grows almost linear in $\Delta$ in the numerically accessible regime
\begin{equation}
\elp \sim \Delta^\gamma, \qquad \gamma\approx 1.1 \pm 0.1 \label{expBulk}
\end{equation}
\n{Below we will argue that the exponent $\gamma$ is not a `true' asymptotic exponent but only an effective parameter describing an extended crossover regime.}
\delete{
In Appendix \ref{errorbars} we give details on the determination and the error of this quantity.
This has to be contrasted with the behavior expected from scaling, which predicts a parametrically much larger screening length of order $\Delta^2$, and thus at least an order of magnitude larger for the parameters investigated by us.}
For values of $r \gtrsim 0.4 \Delta^\gamma$, the screening charge exceeds $-Q_0\approx -2$. This implies that there is a substantial amount of overscreening in the system: on average too  much charge of opposite sign accumulates around each charged dopant.  

\delete{To obtain a scaling collapse, we have rescaled only the $r$ axis but not $Q_s$. This is equivalent to the statement that the scaling relation derived from Eq. (\ref{eq:Deltascaling1A}), $\beta=3 \gamma$, is valid. To understand the importance of overscreening, let us assume for a moment that perfect screening would occur on the length scale $\elp$ implying a fast decay of
$\ccrr$ on scales larger than $\elp$. In this case, one can use Eq.~\eqref{eq:Deltascaling1A}
to obtain that the typical size of a potential fluctuation is of the order $\Delta^{\gamma/2} \ll \Delta$,
much too small to create the puddles necessary for screening, in contrast to the assumption as long as $\gamma<2$.

We therefore conclude that the overscreening is ultimately responsible to build up potential fluctuations of sufficient strength. Using the overscreening mechanism, it apparently becomes possible to gain Coulomb energy by bringing opposite charges close to each other and at the same time maintain sufficiently strong potential fluctuations of the order of  $\Delta$.

Unfortunately it was not possible to reliably extract a second, much larger length scale at which the overscreening crosses over to exact charge
compensation \n{and which may also control the long-distance screening \cite{Bara1984,Lee1999}}. 
This has to occur when the system size is reached (we have simulated only systems with exact charge neutrality). 
For large distance the  noise level is too high, although each curve is averaged over $800-900$ disorder realizations.
From the numerical results, we can not even exclude that this second length scale is set by the system size in our simulations. 
Note, however, that the scaling plot of Fig.~\ref{bulksc} and thus the determination of $\elp$ is not affected by finite size effects.
}

We have checked that other observables, for example the potential correlation function or the typical distance of neutral dopants of different type,
show similar scaling behaviors, see Appendix \ref{potcorrs} for an example.
Most importantly, they all consistently show the importance of the length scale $\elp$ which governs not only screening, but also the length scale on which the dominant short-distance fluctuations of the potential occur.
$\elp$ therefore also determines the distance of puddles of opposite charge.
\delete{ We have not found any evidence of a $\D^2$ length scale in any observable, \n{perhaps related to the problems with finite size effects at long distances discussed above}.}

\section{Screening by metallic surface states}\label{surface}
Topological insulators differ from ordinary insulators or semiconductors because topology enforces the existence of conducting surface states. 
These states are of interest in the context of our discussion, because they provide an extra channel for screening. 
STM measurements of surface states can also be used to obtain quantitative information on the strength and length scale of potential fluctuation at the surface \cite{STM2011}. \n{Most importantly, the suppression of puddles in thin slabs of topological insulators is expected to lead to a substantial reduction of the bulk conductivity and should therefore enhance the quality of devices based on topological insulators substantially. A major goal of this section is therefore to estimate how thin a topological insulator has to be so that puddle formation is effectively suppressed. Note that such a suppression can occur even in the absence of metallic surface states, as has been discussed heuristically by Mitin \cite{Mitin2010} for semiconductor  heterostructures.}

The surface states of a $3d$ topological insulator can, generically, be described by a Dirac equation, and thus have asymptotically a density of states proportional to the doping level.
Their electronic properties can be characterized by the surface doping $\mu^S$ and the effective fine structure constant $\alpha=\tfrac{e^2}{4\pi \varepsilon_{\rm{surf}} \varepsilon_0 \hbar v_F}$, where, in vacuum, 
$\varepsilon_{\rm{surf}} = \tfrac{\varepsilon_{\rm{bulk}}+1}{2} \sim 100$.
Typical values for $\alpha$ in Bi-based topological insulators are in the range of $\a \approx 0.1 \dots 0.2$ (using, e.g., $v_F$ taken from ARPES data \cite{Xia2009}).
In Ref.~\cite{Skinner2013}, Skinner, Chen and Shklovskii develop a detailed analytic theory on how bulk impurity states affect the surface. 
We will instead investigate the question  how the screening from surface states feeds back on bulk properties using some of their results.

If the surface possesses a finite doping, described by a finite chemical surface potential $\mu^S$, it can screen charges on a length scale described by the surface screening length $\ell^S_s \sim v_F/(\alpha |\mu^S|)$. We first consider the limit that $\ell^S_s$ is smaller than the distance of bulk
impurities, $\ell^S_s \lesssim N^{-1/3}$ or, equivalently, $|\mu^S| \gtrsim E_c/\alpha^2$. 
In this case the surface state of the topological insulator acts effectively like a perfect metal. Then, screening of a dopant with charge $q_i$ at distance $z$ from the surface is 
described by positioning a mirror charge with charge $-q_i$ at the same distance on the opposite side of the surface. 
This simple screening mechanism can be implemented in a straightforward way into the model described in Sec.~\ref{model}. \n{To model a thin slab of a topological insulator with two metallic surface states, one formally needs an infinite sequence of mirror charges. As described in the appendix~\ref{mirror}, an accurate and numerically efficient description is obtained by using just a single mirror charge and a linear correction term setting the potential to zero at both surfaces.

Besides its importance for applications of TI materials, the problem of an infinitely large TI slab of finite thickness $L_z$ has also a technical advantage which we will use in the following: the 'bare' potential $\Phi_0(\bm{r})$ arising from randomly placed impurities remains finite for finite $L_z$ even in the thermodynamic limit (while it would diverge in the absence of surface screening). 
This is the potential one obtains in the absence of puddle formation when all donors (acceptors) have charge $+1$ ($-1$). One can easily calculate the distribution $p(\Phi_0(z))$ of this potential by averaging over the position of dopants, here $z$ is the coordinate perpendicular to the surfaces. In the following we will focus our discussion for simplicity on the distribution in the middle of the sample, $z=L_z/2$. Due to the central limit theorem, this initial distribution (before puddle formation) is Gaussian
\begin{align}
p^0_{L_z/2}(\Phi)=\frac{1}{\sigma(L_z) \sqrt{2 \pi}} \exp\!\left[{-\frac{1}{2} \left(\frac{\Phi}{\sigma(L_z)}\right)^2}\right]
\end{align}
where the width of the distribution $\sigma(L_z)$ can simply be computed from $\sqrt{\langle \Phi_0(z=L_z/2)^2 \rangle}$ and, in units of $E_c$, is given by
\begin{align}
\sigma(L_z)&=\left(2 \int_0^{L_z} dz \int_{-\infty}^\infty dx \,dy \,\, V^s(\bm x,(0,0,L_z/2))^2\right)^{1/2} \nonumber \\
&\approx  2.41 \sqrt{L_z} - \frac{7.75}{\sqrt{L_z}}\quad \text{for}\ L_z\gg 1 \label{sigmaAsym}
\end{align}
where the potential $V^s$  is defined in the appendix \ref{mirror} and the second line is a fit to the numerical integral valid for large $L_z$. The prefactor of the leading term depends only on the geometry of the setup but is otherwise universal, the subleading term is linear in the chosen cutoff $a_B$ defined in Eq.~\eqref{eq:Vcutoff}. For small $L_z$, the width $\sigma(L_z)$ is much smaller than the gap, implying that puddle formation cannot take place. As discussed in the introduction, Efros and Shklovskii \cite{Efros1972} estimated the length scale triggering puddle formation (for a different geometry) from the condition $\sigma(L_z) \sim \Delta$, leading to a length scale proportional to $\Delta^2$ as discussed in Eq. \eqref{eq:ESscaling}. We will need in the following a more quantitative version of this argument. We will take into account, that for the successful screening of the potential in the limit $L_z\to \infty$, it is not necessary to redistribute $O(1)$ charges. Instead we can use the total density of neutral dopants in the thermodynamic limit, $n_0$, as an estimate of the volume fraction where the bare potential triggers puddle formation by becoming larger than $\Delta/2$. The 
characteristic width of the slab $L_c^0$ below which puddle formation is suppressed, is therefore estimated from the condition
\begin{align}
p_0(|\Phi|>\Delta/2)&\sim n_0 \label{bareEstimate}
\end{align}
with 
\begin{align}
p_0(|\Phi|>\Delta/2)&=2 \int_{\Delta/2}^\infty p^0_{L_z/2}(\Phi) \, d\Phi
\end{align}
solved by
\begin{align}
\sigma(L_c^0)^2
&\sim \frac{\Delta^2}{8 \ln[1/(n_0 \sqrt{\pi \ln[2/(\pi n_0^2)]/2}]} 
\end{align}
for small $n_0$. Using Eq.~\eqref{sigmaAsym}, we find 
for  $\Delta \to \infty$ as an estimate for the characteristic width $L_c^0$
\begin{align}
L_c^0 \sim \frac{\Delta^2}{46.5 \ln[1/n_0]}\approx  \frac{\Delta^2}{139 \ln[\Delta]}\label{Lcasym}
\end{align}
with (sizable) relative corrections of the order of $\ln[\ln \Delta]/\log[\Delta]$. In the last line we uses that in the asymptotic regime, $n_0 \sim 1/\Delta^3$, \ see Eq.~\eqref{eq:ESscaling}. This formula misses  logarithmic corrections to $n_0$ which give, however only subleading terms beyond the precision of Eq.\eqref{Lcasym}. While our equation can only be an order-of-magnitude estimate, we have kept multiplicative numerical prefactors to indicate their rather large numerical value. 

Note that Eq.~\eqref{bareEstimate} and therefore also \eqref{Lcasym} was obtained only by considering properties of the bare potential, not including any self-consistent screening effects. The formulas can therefore only be viewed as a crude estimate of the relevant length scale of the problem obtained by extrapolating from the bare potential. The result clearly suggests the presence of logarithmic corrections to scaling but we cannot exclude that a resummation of logarithms leads to a modification of the logarithm, e.g., $L_c \sim \Delta^2/\ln^\alpha[\Delta]$. 
A scaling $L_c \sim \Delta^\gamma$ for $\Delta \to \infty$ with an exponent $\gamma$ smaller than $1$, can, however, be excluded as in this case the regions where the bare potential can trigger puddle formation, are exponentially suppressed.
In the following  we will compare the estimate from condition Eq.~\eqref{bareEstimate} to the full numerical solution obtained for moderately large values of $\Delta \lesssim 35$ and find that the formula nevertheless reproduces  the approximately linear $\Delta$ dependence ($\gamma \approx 1.1 \pm 0.1$) in the regime, $\Delta \lesssim 35$.
}

\begin{figure}[t!]
	\centering
		\includegraphics[width=0.9 \columnwidth]{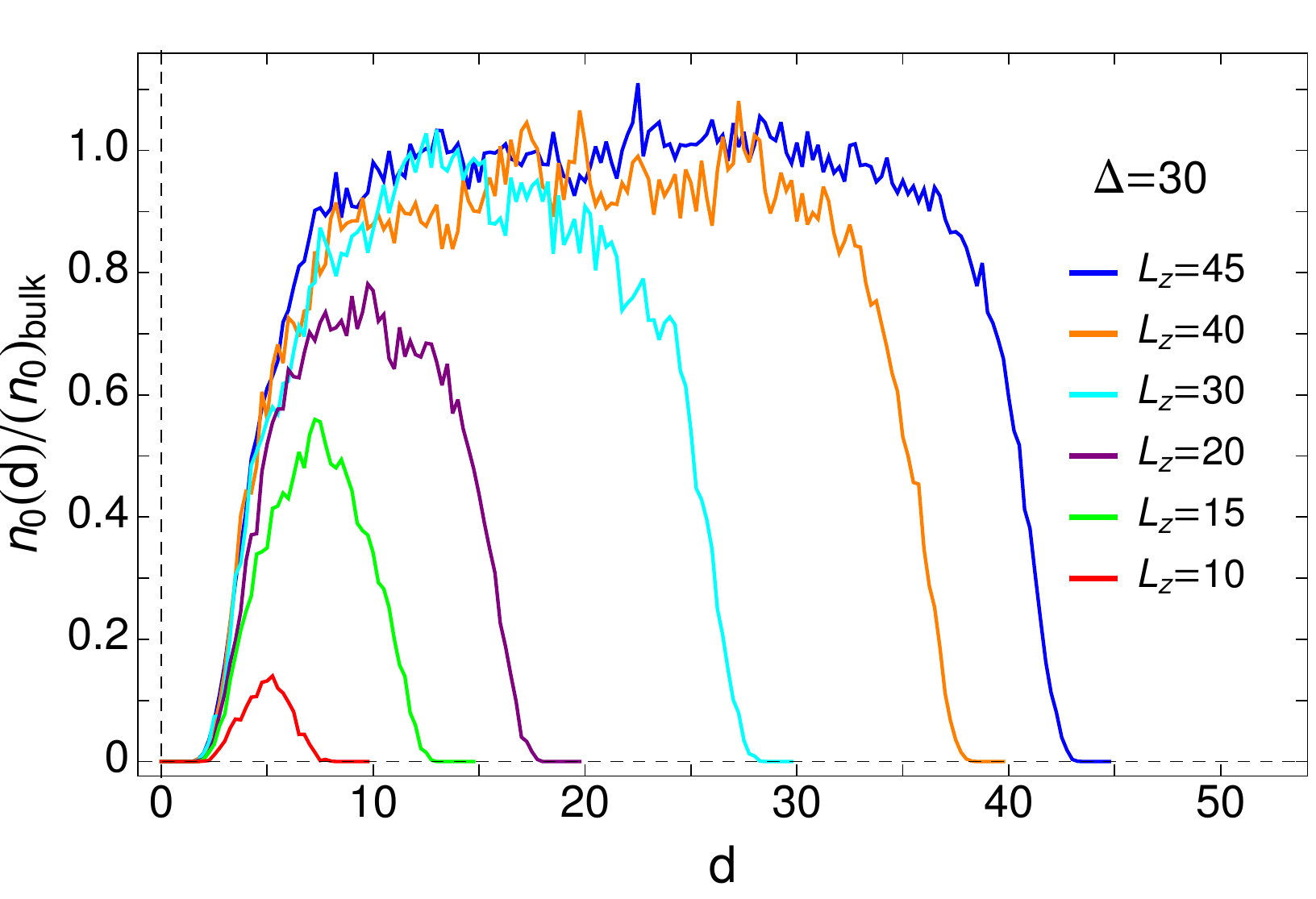}\\
		\includegraphics[width=0.9 \columnwidth]{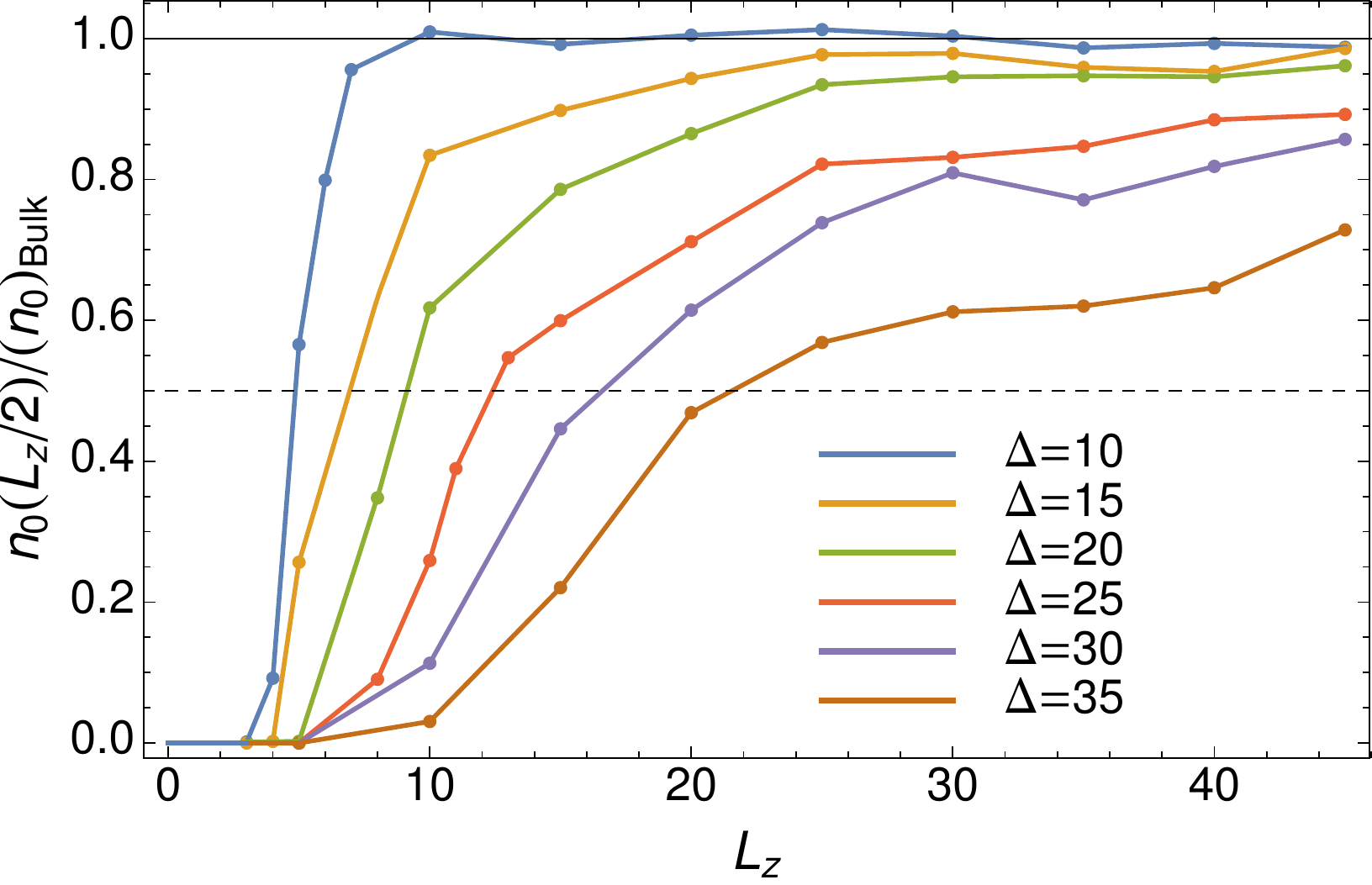}
	\caption{(Color online)  \n{Density of neutral dopants $n_0(d)$ in a slab of a topological insulator of width $L_z$, where $d$ is the distance from one of the surfaces ($\Delta=30$). The metallic surface states screen the potential, thus suppressing $n_0(d)$. For thick samples, $L_z\gtrsim 30$, $n_0(d)$ is only suppressed close to the boundaries while in the center one recovers the bulk puddle density. For $L_z \lesssim 20$, $n_0(d)$ also drops in the middle of the sample  (simulations for $L_{x,y}=50$, $2 \cdot L_x L_y L_z = 5000 L_z$ dopants, averaged over $300$ disorder configurations). The lower panel show the density of defects in the center, $n_0(L_z/2)$, as a function of $L_z$ for various values of $\Delta$.}
	}
	\label{twoSurfaces}
\end{figure}

The surface screening will suppress potential fluctuations and the formation of puddles close to the surface. 
Therefore, all donors will have charge $+1$, all acceptors have charge $-1$ and the density of neutral dopants vanishes close to the surface. 
This physics can be captured by computing the density of neutral dopants, having a charge $0$, as a function of distance from the surface
\begin{equation}
n_0(d)=\expv{\sum_i \delta(d-z_i) \delta_{q_i,0} },
\end{equation}
where $z_i$ is the  (dimensionless) distance of dopant $i$ from the surface and $\delta_{i,j}$ denotes the Kronecker delta. \n{In contrast to charge and potential, the density of neutral dopants is a quantity not fluctuating in sign and therefore its average is both easier to compute (statistical fluctuations are much weaker) and to interpret.

\begin{figure}[t!]
	\centering
		\includegraphics[width=0.9 \columnwidth]{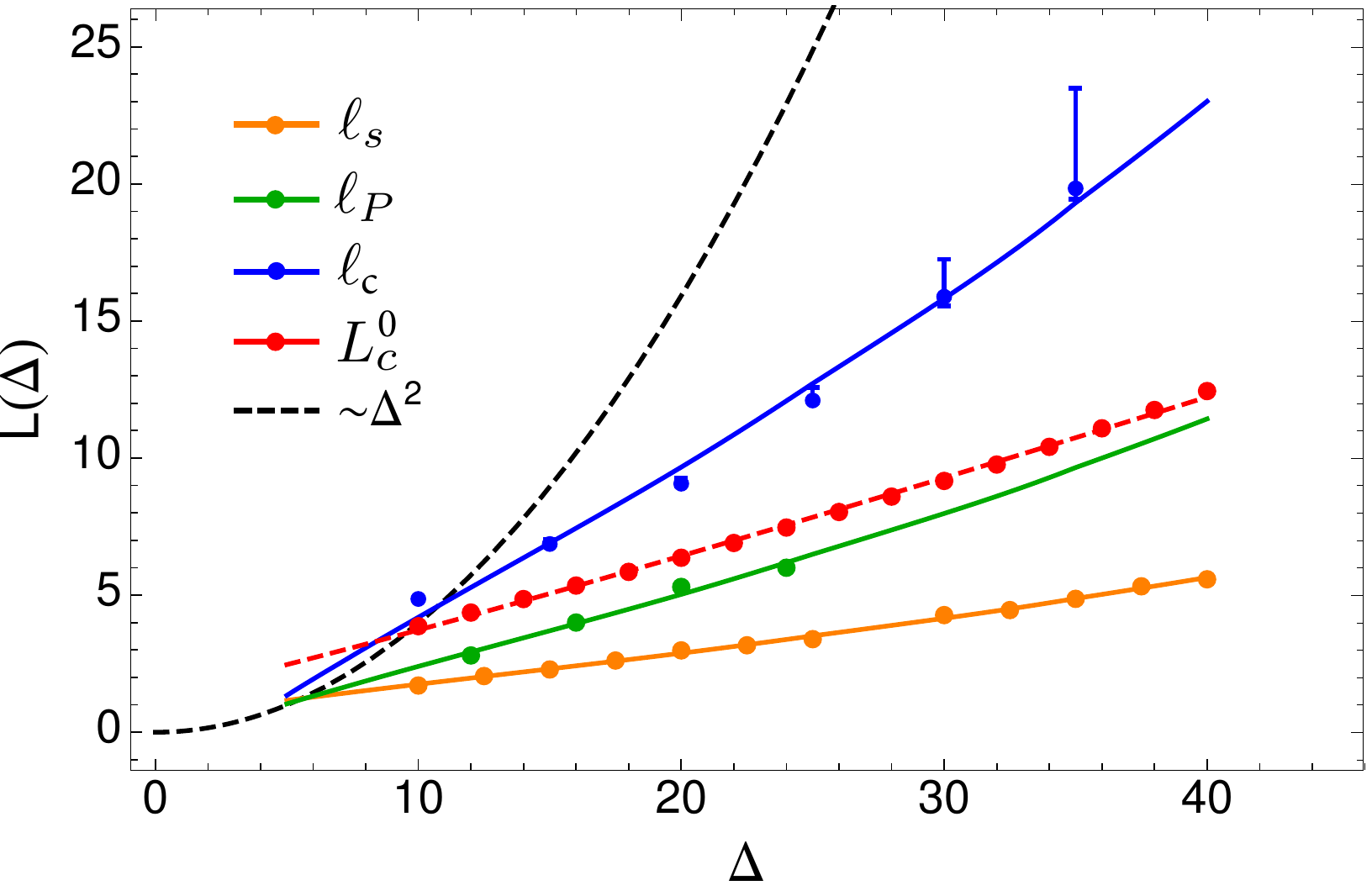}\\
		\includegraphics[width=0.91 \columnwidth]{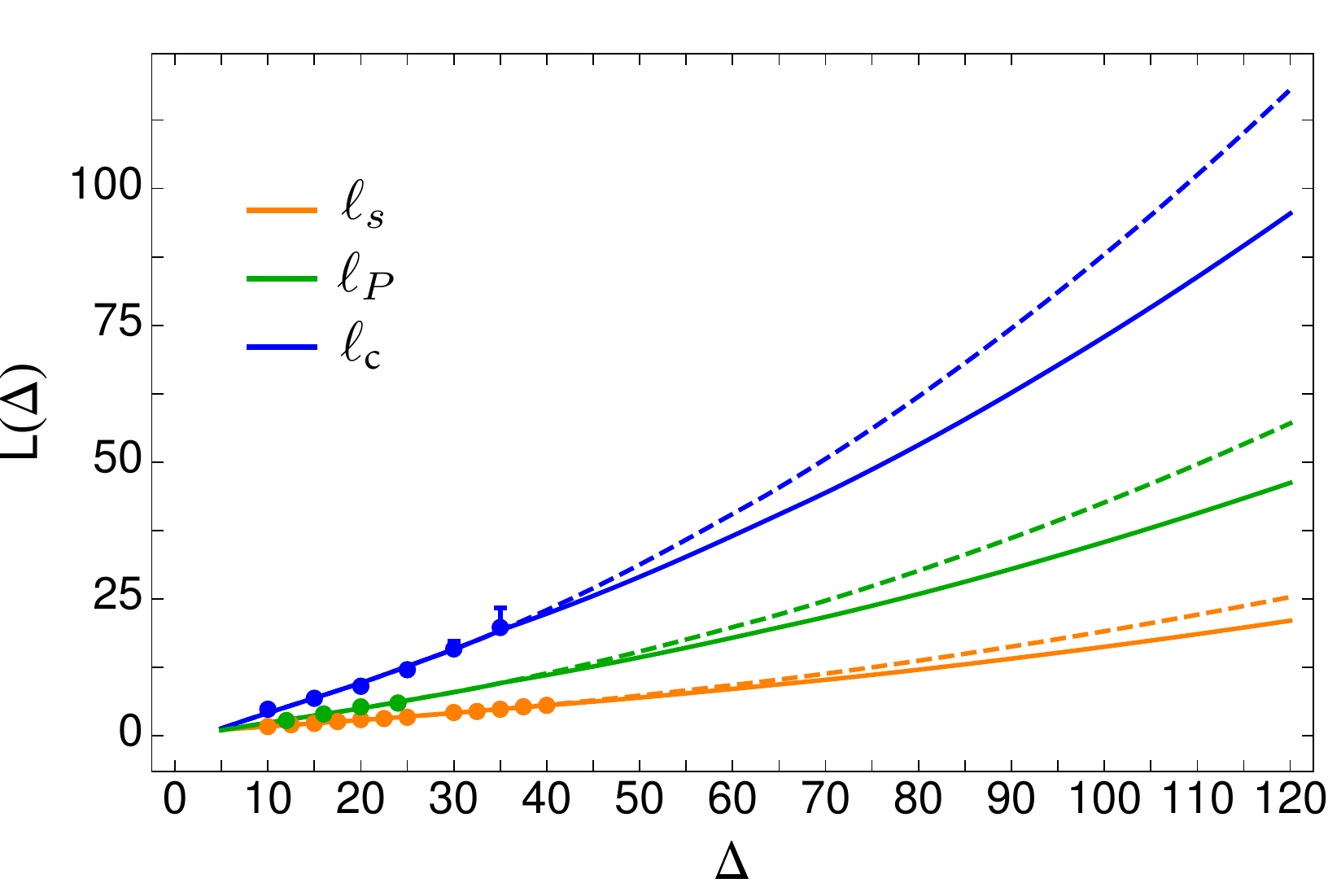}
	\caption{(Color online) \n{$\Delta$ dependence of four different length scales,  $\ell_P, \ell_c, \ell_s$ and $L_c^0$. Three of them, $\ell_P, \ell_c, \ell_s$, have been obtained numerically, see Eq.~(\ref{ls1}--\ref{ls3}).  $\ell_P$ characterizes the screening of charges (on average) in the bulk, $\ell_c$ the suppression of puddles in a thin slab of a topological insulator, and $\ell_s$ the suppression of puddles close to a metallic surface. The error in $\ell_c$ arises from the error in the determination of $n_0$ in the thermodynamics limit, see appendix~\ref{densityNeutral}. $L_c^0$ is an analytic order-of-magnitude estimate for $\l_c$ based on Eq.~\eqref{bareEstimate} (using numerically determinated values for $n_0$).  The black dashed line in the upper panel is the curve $\Delta^2/(8 \pi)$ \cite{Skinner2013Rev} which shows that all length scales rise much slower than $\Delta^2$, the red dashed line is a powerlaw fit $1.36+ 0.19 \Delta^{1.1}$ to $L_c^0$.
	To extrapolate to larger values of $\Delta$, we use a linear fit of $\ell_i$ to $L_c^0$, 
	 $\ell_i =a_i' + c_i' L_c^0$ with $a_i'=-1.67,\ -4.29,\ -0.07$ and $c_i'=1.05,\ 2.19,\ 0.46$ for $i=P,c,s$, respectively (solid lines in both panels). For $\Delta>35$ (lower panel), $L_c^0$ was determined assuming $n_0(\Delta)=n_0(35) (35/\Delta)^3$ (solid lines), see text. The dashed lines, calculated from $n_0(\Delta)=n_0(35) (35/\Delta)^{1.62}$, are shown to indicate how sensitive the result is to a different extrapolation of $n_0$. }
	}
	\label{allScales}
\end{figure}
In Fig.~\ref{twoSurfaces} (upper panel) we show $n_0(d)$ for different values of $L_z$. As expected, the puddle formation and therefore $n_0(d)$ is suppressed close to the two metallic surfaces. For sufficiently thin $L_z$, $n_0(d)$ becomes small even in the center of the sample. The lower panel of  Fig.~\ref{twoSurfaces} therefore shows the density of neutral dopants in the center of the slab, $n_0(L_z/2)$, a function of $L_z$. 

In Fig.~\ref{allScales} we show four different length scales, $\ell_P, \ell_c, \ell_s$ and $L_c^0$, as function of $\Delta$. 
The first three length scales have been extracted from our numerics and the equations
\begin{align}
Q_2(\Delta,\ell_P)&=-1 \label{ls1}\\
n_0(L_z/2) &=\frac{1}{2} n_0^{\rm bulk} \quad \text{for}\ L_z=\ell_c \label{ls2} \\
n_0(\ell_s)&=\frac{1}{2} n_0^{\rm bulk} \quad \text{for}\ L_z \gg \ell_c  \label{ls3}
\end{align}
They describe the characteristic length scale $\ell_P$ on which - on average - a charge is screened in the bulk (see Fig.~\ref{bulksc}), the characteristic width $\ell_c$ of a slab of a topological insulator below which the density of puddles drops to half the bulk value, and the length scale on which puddle formation is suppressed close to the metallic surface of a thick slab of  a topological insulator (see appendix~\ref{singleSurface}).

All three curves show an approximately {\em linear} behavior with $\Delta$ quite different from the $\ell \sim \Delta^2$ (with logarithmic corrections) expected from scaling arguments. All curves are well described by fits of the form  $\ell_i=a_i+c_i \Delta^{1.1}$. Remarkably, the same behavior is also obtained from the  estimate $L_c^0$, which was obtained from the condition in Eq.~\eqref{bareEstimate}, i.e., from properties of the bare potential (before puddle formation). Therefore $\ell_i$ is well described by a linear fit to $L_c^0$ as shown by the solid lines in Fig.~\ref{allScales}.
The dashed red line in Fig. \ref{allScales} shows a power-law fit, $L_c^0\sim \Delta^{1.1}+const.$, which works remarkably well.
We therefore conclude that (i) the average bulk screening and the surface screening are governed by the same length scale, and that (ii) one can use the `naive' scaling argument, \eqref{bareEstimate}, to obtain this length scales (up to multiplicative factors of $O(1)$ and a small offset of $O(1)$).

As we have shown above, the asymptotic behavior for $L_c^0$ is according to Eq.~\eqref{Lcasym} given by $\Delta^2/\ln[\Delta]$ and  definitely not by an exponent close to $1$. The apparent power-law behavior with an exponent close to one therefore reflects only an extended crossover regime: there is no `true' power-law with an exponent smaller than $2$ (see, e.g., Ref.~\cite{reviewQPT} for a discussion on the definition and determination of exponents). The approximately linear behavior for $10\lesssim \Delta \lesssim 50$ arises from the interplay of logarithmic corrections at large $\Delta$ and subleading corrections for small $\Delta$, see Eq.~\eqref{sigmaAsym}.

As also the numerically determined length scales $\ell_i$ show the same behavior, we conclude that also in this case the numerics probes  the same extended crossover regime for numerically accessible vales of $\Delta\lesssim 35$. Below, we will argue that one can use the results for $L_c^0$ to estimate the value of $\ell_i$ for larger values of $\Delta\sim 100$, relevant for Bi-based topological insulators.
}

Above we assumed a perfectly metallic surface state, $|\mu^S| \gtrsim E_c/\alpha^2$, which has a screening length that is short compared to the mean distance of dopants $N^{-1/3}$. Using the result given above, that the screening by bulk states sets in only at a parametrically larger scale, $\ell_s$, we can relax this requirement. Our results should be valid as long as the surface screening length $\ell_S^S \sim v_F/(\alpha |\mu^S|)$ is small compared to $\ell_s$, or
$|\mu^S| \gg v_F/(\alpha \ell_s)$. 

Here $\mu^S$ denotes an effective chemical surface potential. 
Even if the chemical potential of the surface state is {\em exactly} at the Dirac point, $\expv{ \mu^S}=0$, disorder will induce a finite density of states allowing for screening. 
Due to the charged dopants metallic puddles will form on the surface which can, in turn, screen bulk charges. 
To estimate the effect of these surface puddles (not to be confused with puddles in the bulk) we use the results of Ref.~\cite{Skinner2013} (similar results in the context of graphene have, e.g., been obtained in Refs.~\cite{Hwang2007, Adam2007}). 
For the computation of the resulting surface screening length  $\ell_S^S$ for $\expv{ \mu^S}=0$ the authors of Ref.~\cite{Skinner2013} did not take into account any bulk-screening effects, which is justified as long as
$\ell_S^S \ll \ell_s$. Under these conditions, Skinner, Chen and Shklovskii \cite{Skinner2013} found that 
$|\mu^S| \sim E_c/\alpha^{2/3}$ or $\ell_S^S \sim N^{-1/3}/\alpha^{4/3}$. \n{From the condition 
$\ell_S^S \ll \ell_s$, we obtain (using our dimensionless units)
\begin{equation}\label{condSS}
\ell_s > c \left(\frac{1}{\alpha}\right)^{4/3} \qquad \text{for } \ \expv{\mu^S}=0
\end{equation}
where $c\approx 0.6$ according to Ref.~\cite{Skinner2013}, where the authors estimate $\alpha \approx 0.24$ for a Bi-based topological insulator, which  results in the condition $\ell_s >4$
for this class of systems.}

If the condition (\ref{condSS}) is fulfilled, the surface state of a topological insulator provides sufficient screening 
to suppress efficiently the formation of puddles within the distance $\ell_s$.

\section{Discussion and Outlook}

\n{
We have investigated the influence of charge dopants in (topological) insulators, focusing on the case of perfect compensation with equal densities of donors and acceptors. Motivated by the physics of Bi-based topological insulators, we studied the limit where the gap $\Delta$ is large compared to the Coulomb energy $E_c$ of neighboring dopants with $\Delta/E_c \sim 100$ as a typical value \cite{Borgwardt2016}. In our numerical simulations we are not able to  such large values of $\Delta$. Therefore analytical estimates are needed to extrapolate to larger values of $\Delta$.

Our main focus has been the investigation of the length scales governing the formation (and destruction) of puddles. As has been pointed out before in the literature \cite{Bara1984,Lee1999}, due to the long-ranged nature of the Coulomb interaction and the highly non-linear screening effects, there is more than one such length scale. We have, however, found that  the size of a (average) screening cloud around an impurity, the typical distance of electron- and hole puddles, and - most importantly - the length scales governing the suppression of puddle formation in the bulk due to metallic surface states, are all similar and show an approximately linear increase with $\Delta$ for $\Delta \lesssim 35$ (a fit gives $\ell_i \sim \Delta^{1.1}$). We have found a simple analytic estimate of such length scales based on properties of the bare potential, which reproduces this behavior in an extended crossover regime but predicts $\ell_i \sim \Delta^2/\ln[\Delta]$ for $\Delta \to \infty$. 

One can use this analytic estimate to obtain a quantitative extrapolation of the numerically determined results to larger values of $\Delta$. This is shown in the lower panel of Fig.~\ref{allScales}. The fit $\ell_i = a'_i + c'_i L_c^0$ gives an excellent fit to the numerically determined length scales $\ell_i$ (see figure caption for details and fit parameters). Using this extrapolation, we can estimate the corresponding length scale for large values of $\Delta$.

Assuming, for example, $\Delta/E_c \sim 100$ and $N\approx 10^{19}\,{\rm cm}^{-3}$, our best estimates for the dimensionless length scales are $\ell_c\approx 72.9 \pm 15.0$, $\ell_s \approx 16.3\pm 2.8$ where errors have been estimated based on the use of different extrapolations of $n_0$, see Fig.~\ref{allScales}. In physical units this implies that the width of the region close to the metallic surface where puddle formation is inhibited is about $62.6-88.6\,$nm. Puddle formation in the center of a thin slab of a topological insulator is predicted to be suppressed by metallic surface states by at least a factor 2 if the slab is thinner than $268.7-407.9\,$nm. As puddles largely control the bulk conduction at low temperatures by reducing the energy gap for transport processes \cite{ESbook,Skinner2012, Skinner2013Rev}, the suppression of puddle formation in the bulk is expected to be accompanied by a strong suppression of bulk conduction. More precisely, at least three effects will contribute to the increase of bulk resistivity the screening from metallic surfaces, the suppression of Coulomb fluctuations due to the dimensional crossover (even without metallic surfaces), and the crossover from a 3d to a 2d percolation problem of electrons moving in a correlated potential \cite{ESbook,Mitin2010}. As we have shown that surface and bulk effects are governed by similar length scales proportional to each other, we expect that all effects are of similar importance.
In compensation doped Bi-based compounds the bulk conductance is expected to be suppressed considerably (i.e. much faster than to be expected from geometric factors)
when the slab becomes thinner than, e.g., $270$\,nm. It will be interesting to develop a quantitative theory for transport in the future which combines numerical calculations for smaller values of 
$\Delta/E_c$ with analytic extrapolation schemes for large $\Delta$ similiar to the ones used in this paper.

}

\section*{Acknowledgments}
We thank  Y.~Ando, O.~Breuning, N.~Borgwardt, M.~Gr{\"u}ninger, \n{B. Skinner, and, especially,
B. Shklovskii} for insightful discussions and useful comments.
The numerical simulations have been performed on the CHEOPS cluster at RRZK Cologne. We thank the DFG for financial support within project C02 of CRC 1238.

\appendix

\section{Sum rules and scaling arguments}

From the definition of $\ccrr$, Eq.~\eqref{eq:rhorhoparam}, and the Poisson equation one can derive a set of exact sum-rules
\begin{align}
\frac{\expv{H_{\rm C}}}{V} &=  \phantom{-}2 \pi \int ds \, s^{1} \,\ccrr(\D,s), \label{eq:hiE} \\
\frac{\expv{H_{\rm n}}}{V}= n_0  N \Delta& = \phantom{-}2 \pi \, \D \int ds \, s^2  \,\ccrr(\D, s) + \D, \label{eq:hiEn0} \\
Q_0=\frac{\expv{Q}}{V} &=  -4 \pi \int ds \, s^2 \,\ccrr(\D,s), \label{eq:hiQ} \\
\expv{\phi^2} &= -8 \pi^2 \int ds \, s^3 \,\ccrr(\D,s). \label{eq:hiff}
\end{align}
Here $\expv{H_{\rm C}}$ is the disorder average of the Coulomb energy, $\expv{H_{\rm n}}$  are the single particle energies of the dopants, see Eq.~\eqref{eq:ham_units}, $\expv{Q}$ is the number of
ionized dopants (not counting the neutral ones), and $\expv{\phi^2}$ is the expectation value of the square of the potential (all expressed in our dimensionless units). 

We can use these sum-rules to obtain a more rigorous version of the scaling argument given above.
We start from the assumption that the physics of the system is governed by a single length scale \n{large compared to the average distance of impurities}.
In this case  $\ccrr(\D,s)$ can be written as 
\begin{equation}
\ccrr(\D,s) = \D^{-\beta} \, \overline{\ccrr} (s/\D^{\gamma}).
\end{equation}
We will show that from this assumption alone Eq.~\eqref{eq:ESscaling} can be derived. Later, we will conclude that the scaling ansatz is {\em not} fully valid: there are substantial subleading corrections even for values of $\Delta\lesssim 35$ and logarithmic corrections in the $\Delta \to \infty$ limit. In this appendix, we will, however, explore only the consequences of the scaling ansatz.

The bulk fluctuations of the potential are of the order of the bandgap $\Delta$, see Fig.~\ref{potfluct}, which implies $\expv{\phi^2} \sim \D^2$.
Furthermore, as the fraction of neutral atoms vanishes for large $\Delta$,  $n_0 \to 0$ for $\D/\Ec \to \infty$,  the density of charged dopants, $Q_0$ is of order $\Delta^0$ with only subleading corrections.
To leading order we therefore obtain from Eqns.~(\ref{eq:hiQ}) and (\ref{eq:hiff}) 
\begin{align}
 \D^0 & \sim \int ds \,s^2 \, \ccrr(\D,s) =
  \D^{-\beta+3 \gamma} \int ds \,s^2 \, \overline{\ccrr}(s),  \label{eq:Deltascaling1A} \\
 \D^2 & \sim \int ds \,s^3 \, \ccrr(\D,s) =
  \D^{-\beta +4 \gamma} \int ds \,s^3 \, \overline{\ccrr}(s) \label{eq:Deltascaling2A}.
\end{align}
Therefore, the scaling ansatz predicts $\beta=3 \gamma$ and $2=-\beta+4 \gamma$ or, equivalently, $\gamma=2$ and $\b=6$, implying a length scale $ \sim \D^2$ and a typical charge density $\sim 1/\D^3$.
This is just a refined version of the argument presented above in Eq.~(\ref{eq:ESscaling}).

From Eq.~(\ref{eq:hiE}) and Eq.~(\ref{eq:Deltascaling1A}) we further deduce that the Coulomb energy density $\expv{H_{\rm C}}/V \sim - \D^{-\beta+2 \gamma}= - \D^{- \gamma}$ ( $\ccrr$ is negative as it describes screening, for example the accumulation of negative charge around a positive one).
This implies that the Coulomb energy is minimized by choosing the screening length $ R_s \sim \D^\gamma $ as small as possible. 
However, this minimization competes with the increasing $H_{\rm n} \sim n_0 N  \D$, see Eq.~(\ref{eq:hiEn0}), and of
course has to respect the constraints, in particular Eq.(\ref{eq:Deltascaling2A}).

Our numerical result are in strong disagreement with the scaling result. Several factors play a role: First, even for $\Delta \sim 35$ the asymptotic scaling regime is not yet reached. It can be seen from Fig.~\ref{allScales}, indicating that both $\ell_s$ and $\ell_P$ are well below $10$ in this regime. Second, a more quantitative estimate based on properties of the bare potential strongly suggested the presence of logarithmic corrections, see Eq.~\ref{Lcasym}. This may indicate that in Eq.~(\ref{eq:Deltascaling2A}), $\int ds \,s^3 \, \ccrr(\D,s)$ obtains logarithmic corrections from a slow decay $\sim 1/s^4$ of  $\ccrr(\D,s)$.

\section{Correlations of the potential} \label{potcorrs}
\begin{figure}[t!]
	\centering
		\includegraphics[width=\columnwidth]{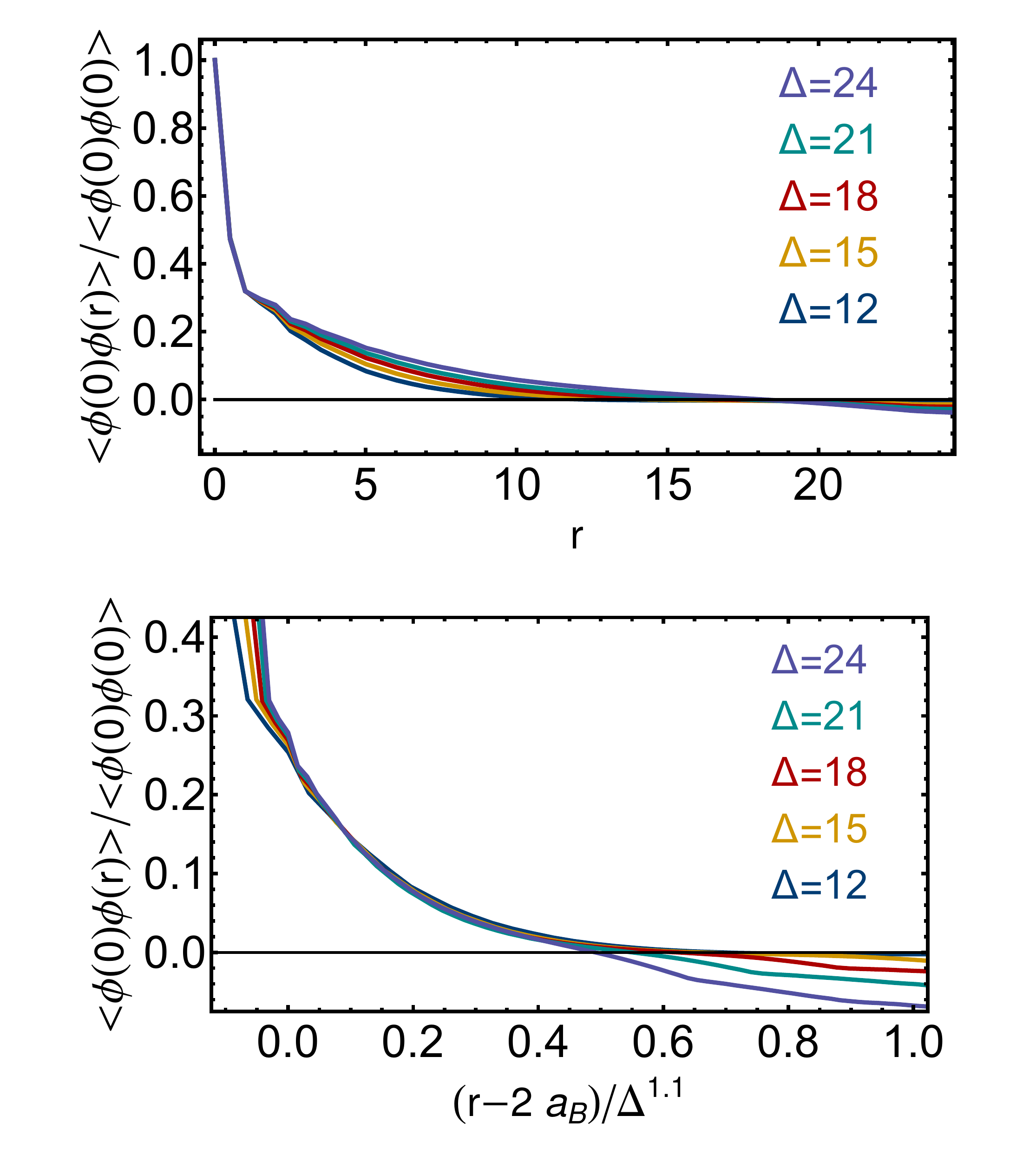}
	\caption{(Color online)  The potential correlation function $\expv{\phi(\br) \phi(0)}$ normalized to $\expv{\phi(0) \phi(0)}$ allows to determine the length scale of fluctuations of the potential.
	Upper panel: Unscaled data. Lower panel: Scaling plot for $\Delta =12...24$. 
	For the scaling of the horizontal axis, we first subtract a short-distance cutoff (see text) and then use the scaling of the bulk screening length $\elp \sim \D^{\gamma}$ where $\gamma \approx 1.1$, see 
	Eq.~(\ref{expBulk}), see text. Scaling breaks down both at short distances of the order of the cutoff and for larger distances, likely related to a second, longer length scale related to overscreening.}
	\label{phi2}
\end{figure}

As show in Fig. \ref{potfluct} the potential in the bulk of a compensation-doped insulator 
fluctuates in space. It is approximately restricted to the range $[-\Delta/2,\Delta/2]$ and exceeds $\pm \Delta/2$ by an amount of order $E_c$ only in the 
region where puddles form. The correlation function $\expv{\phi(r)\phi(0)}$ shows on which length scale the characteristic potential fluctuations occur.
                                                                                                                                                       
In Fig.~\ref{phi2} we show  $\expv{\phi(r)\phi(0)}$ normalized to $\expv{\phi(0)\phi(0)} \sim \Delta^2$. 
At short distances (of the order of the distance of impurities), this is governed by the autocorrelation of the potential of a single charge and decays on a length scale set by $a_B$.
As can be seen in the upper panel of Fig.~\ref{phi2}, the normalized correlation function is independent of $\Delta$ in this regime. The next-largest length scale, the screening length,
on which the correlations decay is more interesting. As expected, we find that this length scale is governed by the bulk screening length  $\elp \sim \Delta^\gamma$, see Eq.~\ref{expBulk}.
This is shown by the scaling plot in the lower panel of Fig.~\ref{phi2}. Clearly the same length scale $\approx 0.2 N^{-1/3} (\Delta/E_c)^{1.1}$ (including prefactors) determines the screening radius 
and the dominant length scale of potential fluctuations.
Note that scaling does not hold at the short distances ($\lesssim a_B$ and/or impurity distance $N^{-1/3}$) and that we had to subtract a short distance cutoff
to obtain a reasonable scaling collapse.  

At larger length scales, the correlation function becomes negative. This physics is, however, {\em not} governed by $\elp$ as follows from
the absence of a scaling collapse in this regime. As discussed in the main text, the physics in the second regime is related to overscreening and 
occurs on a length scale which we cannot resolve with our numerical simulations.

\section{Density of neutral dopants}\label{densityNeutral}
\n{ 
\begin{figure}[t!]
	\centering
		\includegraphics[width=0.8 \columnwidth]{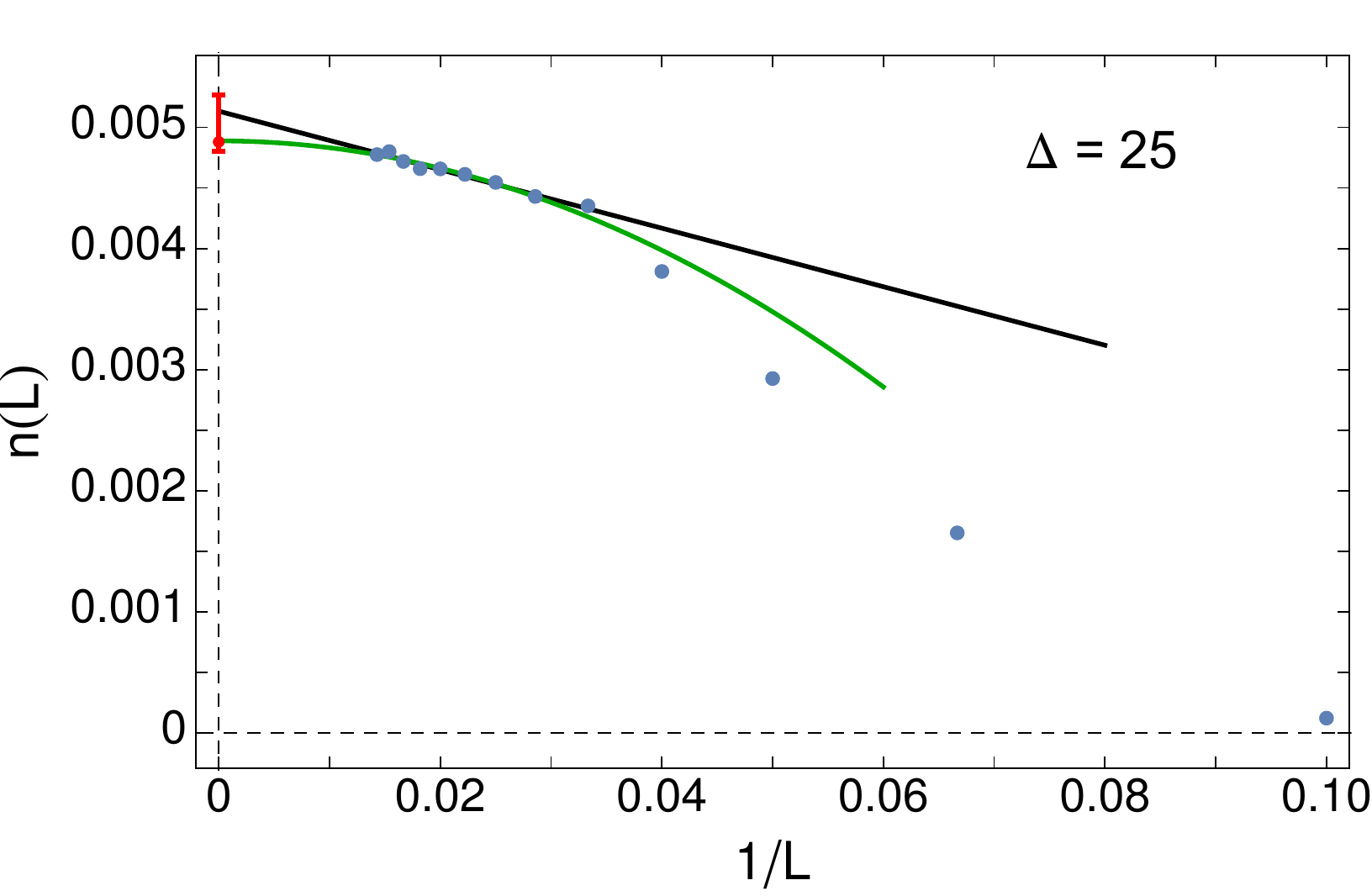}
		\includegraphics[width=0.8 \columnwidth]{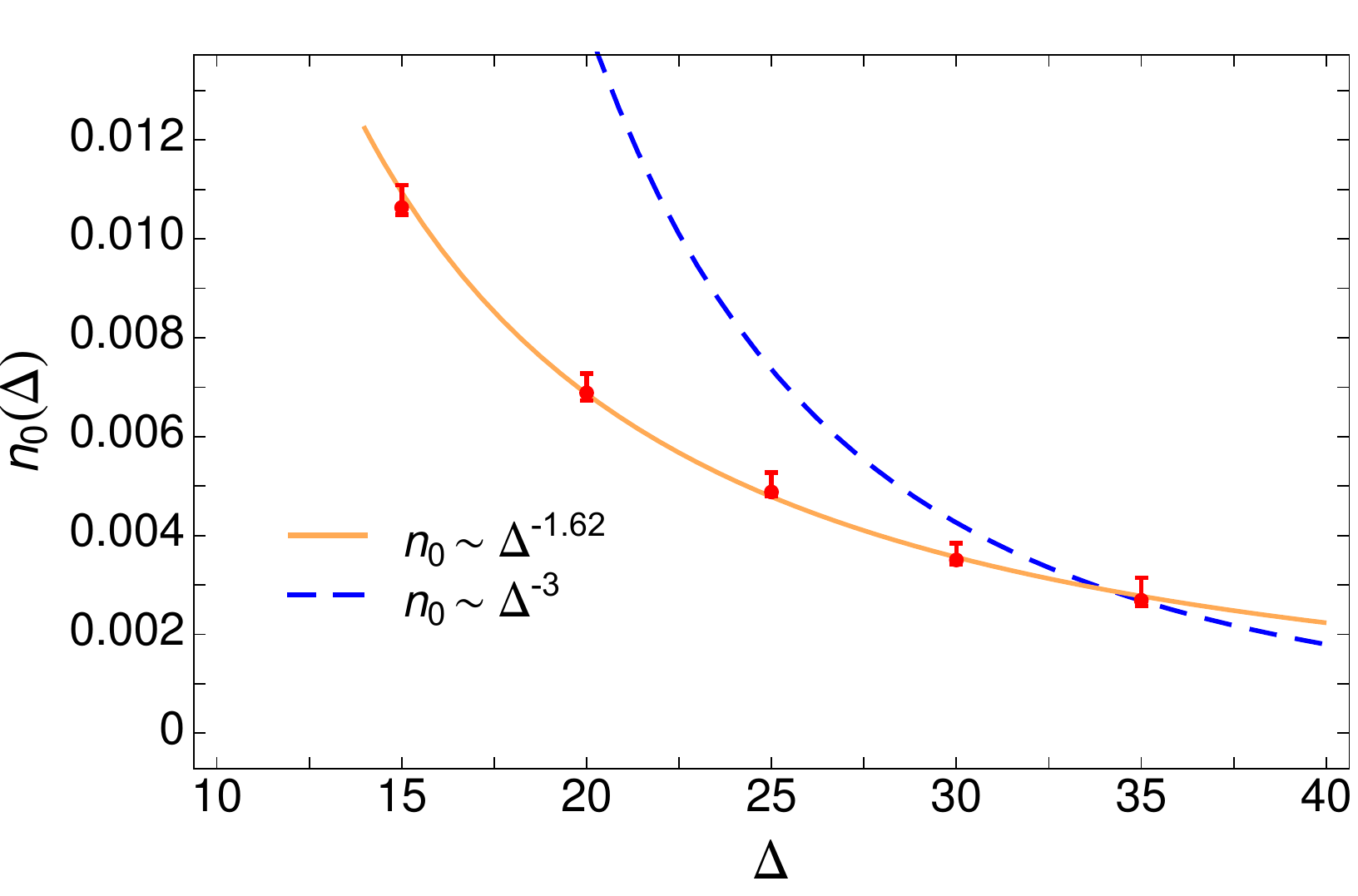}
	\caption{(Color online) \n{\label{figDensity} Upper panel: density of neutral dopants obtained from simulation of boxes of size $L^3$ with periodic boundary conditions. Largest simulations include $2 \times 70^3\sim 700.000$ dopants. To extrapolate to $L\to \infty$ we use both a quadratic (green) and linear (black) extrapolation schemes, assuming finite size errors of order $1/L^2$ and $1/L$, respectively. Error bars (red) are determined by combining the $1 \sigma$ error intervals of both schemes. Lower panel: $\Delta$ dependence of the density of neutral dopants in the thermodynamic limit. For 
	the range of $\Delta$ accessible to our numerics, the data is consistent with $1/\Delta^{1.6}$ (orange line) reflecting an extended crossover regime. A
	 $1/\Delta^3$ dependence  (dashed line)  does not fit the data for $\Delta \lesssim 35$.}}
\end{figure}
To determine the density of neutral dopants in the thermodynamic limit, we have simulated boxes of size $L^3$ with periodic boundary conditions. The resulting density for $\Delta=25$ is shown in the upper panel of Fig.~\ref{figDensity} as function of $1/L$.  As we do not know the analytic $1/L$ dependence of that quantity and as the numerical result is consistent with different interpolating functions, we use the following procedure. To estimate the density in the thermodynamics limit, we use a quadratic extrapolation scheme (green line), assuming that finite size effects are of order $1/L^2$. Within the statistical error bars this is approximately equivalent to the value obtained for the largest system size used in our numerics. A linear extrapolation in $1/L$ (black line) gives a higher value for $n_0(L\to \infty)$. The error bar is obtained from the largest and smallest one-standard-deviation values obtained from both interpolation schemes. It therefore reflects not only the statistical uncertainty of our data but also the much larger systematic error related to the unknown $L$ dependence of finite size effects.

The lower panel of Fig.~\ref{figDensity} shows how the density of neutral dopants drops for increasing $\Delta$. In the crossover regime accessible to our numerics, $n_0(\Delta)$ decays much slower than the $1/\Delta^3$ law expected up to logarithmic corrections from the scaling arguments, see Eq.~\eqref{eq:ESscaling}. This slow decay is, however, consistent with the slow rise of the length scales characterizing screening, which increase approximately linear instead of quadratically with $\Delta$ in the same parameter regime. 
}

\section{Screened Coulomb potential in the presence of two metallic layers}\label{mirror}
\n{
The effective potential of a single impurity with (dimensionless) coordinate $\bm{x}_i$ with distance $x^3_i$ from a metallic layer at $x^3=0$ is described by
(using the same conventions and cutoffs as in Eq.~\eqref{eq:Vcutoff}) 
\begin{align}
V^m({\bm{x}_i},\bm{x})= V[\bm{x}-(x^1_i,x^2_i,x^3_i)]
-V[\bm{ x}-(x^1_i,x^2_i,-x^3_i)]
\end{align}
A "mirror charge" guarantess that the potential vanishes on the metallic surface.
In the presence of two metallic surfaces located at $z=0$ and $z=L_z$, an infinite sequence of mirror charges is required. For efficient simulations it is mandatory that the screened potential can be computed rapidly. We have found that a potential containing only the mirror charge to the closest metallic surface in combination with a linear correction term which sets the potential to zero at both surfaces is sufficiently accurate and numerically efficient, see Fig.~\ref{figPot}. We therefore use in our simulations
for $z_i \le L_z/2$
\begin{figure}[t!]
	\centering
		\includegraphics[width=0.8 \columnwidth]{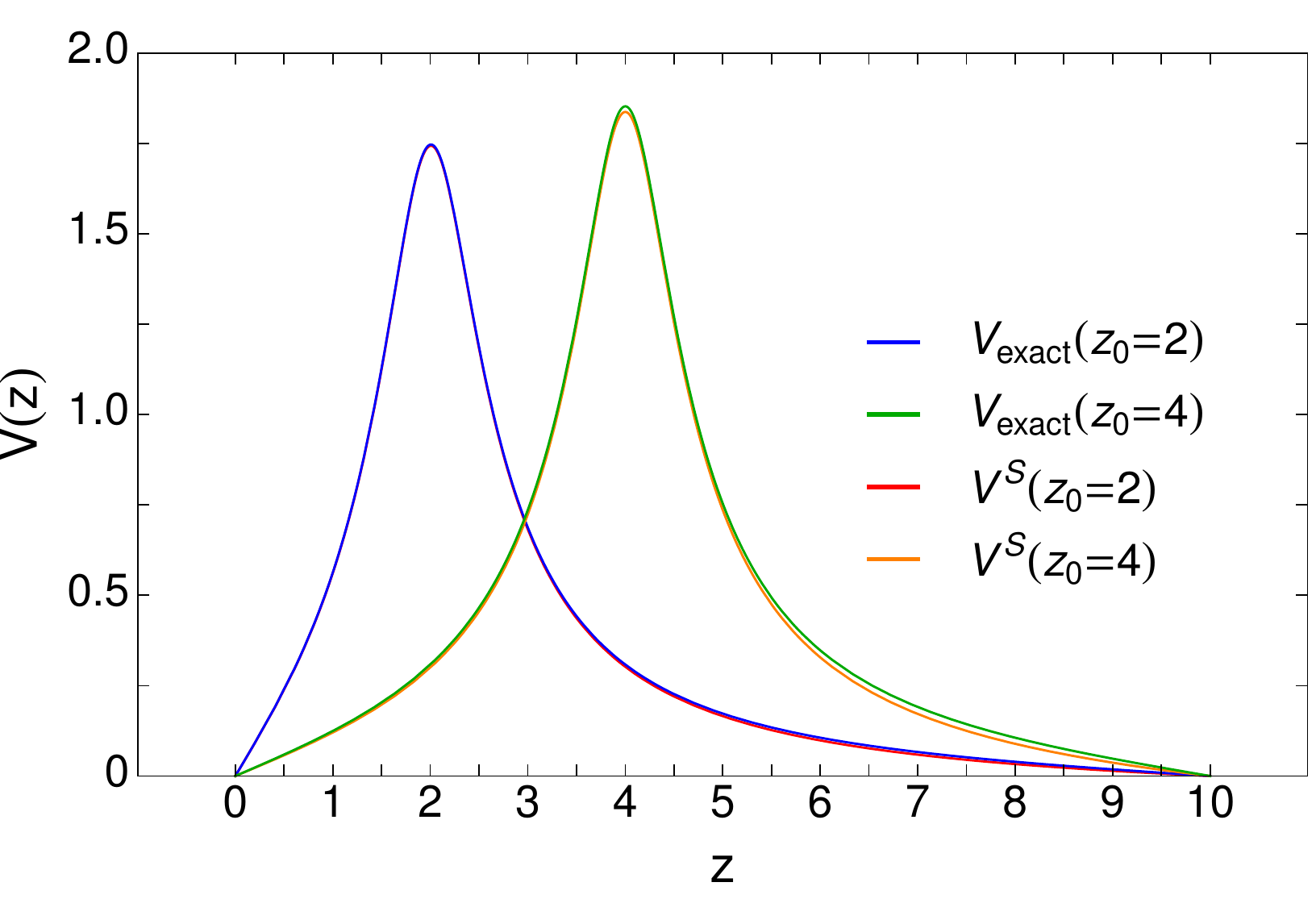}
	\caption{(Color online) \label{figPot}
	Potential of a charged impurity located at $z=2$ and at $z=4$ in the presence of two metallic surfaces at $z=0$ and $z=10$. The plot compares the exact result (upper blue and green curves) to the approximation given by  Eq.~\ref{screenedV} (lower red and orange curves).}
	\label{chi2}
\end{figure}
\begin{align}\label{screenedV}
V^s({\bm{x}_i},\bm{x})&= V^m(\bm{x}_i,\bm{x}) -\dfrac{z}{L_z} V^m(\bm{x}_i,(x_1,x_2,L_z))
\end{align}
For  $z_i >L_z/2$ we use a mirror image of the potential given above with mirror plane $z=L/2$. Due to this construction the derivative of the potential with respect to $z_i$ has a small jump at $z_i=L_z/2$. We have found that this leads to a tiny, hardly visible bump at $z=L_z/2$ in the density of neutral dopants, $n_0(z)$. When we determining the density of neutral dopants  in the center of the slab, we fit a parabola to $n_0(z)$  for $0.4 L_z < z < 0.6 L_z$ omitting a tiny region of width $0.04 L_z$ around $z=L_z/2$. In practice, the tiny bump (and the small correctiont to the fit described above) does, however, have 
no qualitative or quantitative influence on our results.
}

\delete{
\section{Determination of  $\gamma$}\label{errorbars}
\begin{figure}[t!]
	\centering
		\includegraphics[width=\columnwidth]{chi2plot.pdf}
	\caption{(Color online) 
	The quality of the scaling plots is determined from the square deviation $\chi^2(\gamma)$ defined in Eq.~\eqref{x2}. 
	$\chi^2(\gamma)$ is shown as a function of $\gamma$ in red and gray for the scaling plot of Fig.~\ref{bulksc} and Fig.~\ref{surfsc}, respectively.
	For both quantities $\gamma_{\rm min} \approx 1.1$ is obtained as the best exponent given by the minimum of $\chi^2$. See the text for a discussion of the error bars.}
	\label{chi2}
\end{figure}

The exponent $\gamma$ governing the screening length $\elp$ as function of $\Delta/E_c$ has been determined in our paper from both the $r$ dependence of the screening charge $Q_s(r)$, see Fig.~\ref{bulksc},
and from the density of neutral impurities close to a metallic surface, see Fig.~\ref{surfsc}. The exponent can either be obtained from fits (on a log-log scale) of the half-width as function of $\Delta$, see lower panel 
of Fig.~\ref{surfsc}, or from the quality of the scaling collapse.
To obtain a quantitative value for the latter, we calculate for a set of scaling functions $C_i$ the square deviation $\chi^2(\gamma)$ defined by
\begin{eqnarray}
 \chi^2(\gamma) &=& \frac{1}{2}\sum_{i,j} \xi_{ij}^2 (\gamma) \label{x2}, \\
\xi_{ij}^2 (\gamma) &=& \int_0^{r_{\rm max}} \, dr\, |C_i(r/\D_i^\gamma)-C_j(r/\D_j^\gamma)|^2. \nonumber
\end{eqnarray}
$r_{\rm max}$ was chosen such that all points are in the relevant scaling regime.
The resulting $\chi^2$ is shown for both, the screening charge and the suppression of puddle formation at the surface, in Fig.\ref{chi2}.
From all methods we obtain as a best estimate for the exponent $\gamma \approx 1.1$.

The error bars are more difficult to estimate as they are mainly determined by systematic and not by statistical errors related to averaging over disorder configurations.
A main source of systematic errors is that the screening length $\elp \approx 0.2 (\Delta/E_c)^{1.1}$ exceeds the average distance of impurities ($=1$ in our units) only by a factor
of 7 for the largest value of $\Delta$ used in our simulations. We do not have a reliable method to estimate this systematic error. From the width of the minimum in $\chi^2$ we  
estimate the error in $\gamma$ from $\elp$ to $0.1$ and from $w_s$ to $0.2$. Importantly, the value $\gamma=1$ is consistent with our numerical results.
}

\section{Screening from a single surface}\label{singleSurface}

\begin{figure}[t!]
	\centering
		\includegraphics[width=\columnwidth]{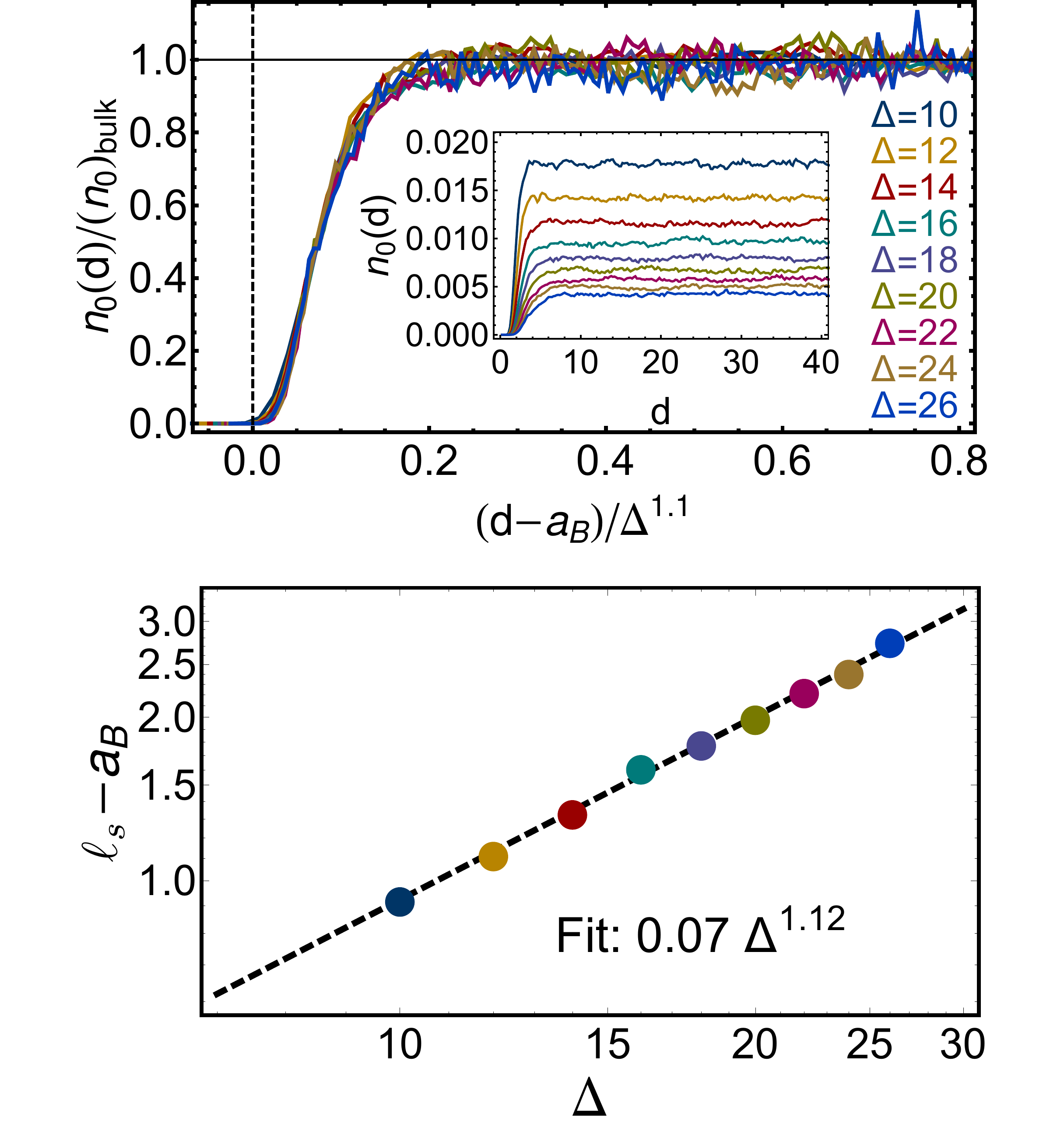}
	\caption{(Color online) Screening from  surface-states of a topological insulator suppresses
	puddle formation close to the surface. Upper panel: Scaling plot of the density of neutral dopants, $n_0(d)$, as a function of the distance $d$ to a metallic surface state ($|\mu^S| \gg E_c/\alpha^2$, $L=50$). Curves are shifted by $a_B=1$ since neutralization starts only for $d>a_B$. 
The inset shows the unscaled data. Lower panel:
The width $\ell_s$ of the surface layer without puddles, defined by $n_0(\ell_s)=(n_0)_{\rm bulk}/2$ as a function of $\D$. A fit gives an exponent $1.12$ very close to the bulk fit, see Eq.~(\ref{expBulk}).
	}
	\label{surfsc}
\end{figure}

In this appendix we briefly discuss the suppression of puddle formation close to an metallic surface (for a system much  thicker than $\ell_c$).
The inset of  Fig.~\ref{surfsc} shows $n_0(d)$ for values of $\Delta$ ranging from 10 to 26. On a relatively short length scale, the bulk value of $n_0(d)$ is reached. 
In the lower panel of Fig.~\ref{surfsc} we plot the width $\ell_s$ of the zone, where surface screening suppresses puddle formation, defined by $n_0(\ell_s)=(n_0)_{\rm bulk}/2$.
After subtracting the offset $a_B=1$ we obtain numerically an approximate power law relation in the numerically accessible regime
\begin{equation} \label{expws}
\ell_s \sim \Delta^\gamma, \qquad \gamma\approx 1.1 \pm 0.2
\end{equation}
We also performed simulation with several other values of $a_B$ ($a_B=0, 0.5, 1.5,2$) and have checked that subtracting $a_B$ results in the same curve (for a fixed value of $\Delta$).
Also the scaling plot in the upper panel of Fig.~\ref{surfsc} confirms that $\ell_d$ governs the size of `dead zone', where puddle formation is suppressed. 

\bibliography{bib_disTI}

\end{document}